\documentclass[11pt]{article}

\usepackage[left=2cm, right=2cm, top=2cm, bottom=2cm]{geometry}
\usepackage{setspace}
\usepackage{indentfirst}
\usepackage{tabularx}
\usepackage{xcolor}
\usepackage{soul}
\usepackage{booktabs}        
\usepackage{threeparttable}
\usepackage{natbib}

\sethlcolor{cyan!20} 

\usepackage[T1]{fontenc}
\usepackage{graphicx}
\usepackage{titlesec}
\usepackage{titling}

\usepackage{amsmath}
\usepackage{amssymb}
\usepackage{booktabs}
\usepackage{makecell}
\usepackage{multirow}
\usepackage{longtable}
\usepackage{array}
\usepackage{enumitem}

\usepackage[hyphens]{url}
\usepackage{hyperref}
\hypersetup{
    colorlinks=true,
    linkcolor=blue,
    urlcolor=blue,
    breaklinks=true
}

\pretitle{%
  \begin{center}%
  \rule{\linewidth}{0.6pt}\\[1.5em]
  \LARGE\bfseries
}
\posttitle{%
  \\[0.75em]
  \rule{\linewidth}{0.6pt}
  \end{center}
  \vspace{1.5em}
}

\preauthor{\begin{center}\large}
\postauthor{\end{center}\vspace{0pt}}

\predate{}
\postdate{}
\date{}  

\begin{document}

\title{Temporal Frictions and Judicial Outcomes:\\[0.5em]
\Large Analyzing the Impact of Time Delays on Criminal Sentencing in Cook County (2020–2024)}

\author{
\large Yifei Tong\\[0.25em]
\normalsize Georgetown University\\[0.25em]
\normalsize\texttt{yt583@georgetown.edu}
}

\maketitle

% Your text starts here: the first paragraph after each section
% will be indented and sit directly under the heading.
\vspace{-15pt} %
\begin{abstract}
This study examines how time delays between criminal offenses and arrests are associated with sentencing outcomes in Cook County, Illinois, during the COVID-19 era. Using administrative court records from 2020 to 2024, the analysis focuses on cases in which arrests did not occur immediately, allowing for systematic variation in procedural delay. The study asks whether longer delays are linked to more severe punishments and whether these associations differ across offense types and institutional contexts during periods of court disruption.

The findings indicate that longer delays are consistently associated with harsher sentencing outcomes, even after accounting for demographic characteristics, case complexity, offense category, and pandemic-related disruptions. These associations are particularly pronounced in violent and sexual exploitation cases. While the analysis does not establish causal effects, the consistency of results across multiple empirical approaches suggests that procedural timing is a meaningful feature of judicial decision-making rather than a neutral administrative artifact. By documenting how institutional delays correlate with punishment severity, this study contributes to empirical research on judicial discretion, court efficiency, and inequality in the administration of justice, highlighting the importance of procedural fairness alongside formal legal criteria.
\end{abstract}

\noindent
{\footnotesize \textit{\textbf{Keywords:}} Criminal Sentencing \textbar{} Sentencing Severity \textbar{} Judicial Decision-Making \textbar{} Temporal Justice \textbar{} Bi-Secting K-Means Clustering \textbar{} COVID-19 Pandemic \textbar{} Cook County Courts \textbar{} Random Forest Classification \textbar{} OLS Regression \textbar{} Regression Modeling \textbar{} Feature Importance \textbar{} Empirical Legal Analytics \textbar{} Institutional Inefficiency}

\section{Introduction}  
\vspace{-5pt} %

The COVID-19 pandemic introduced unprecedented strain on global court systems, slowing case processing and amplifying inequalities in access to justice. In the United States, courts’ reliance on in-person procedures compounded these disruptions, leading to extensive delays and administrative backlogs \citep{viglione2023, shymansky}. As Shymansky argues, pandemic-induced suspensions of trial activity posed a fundamental constitutional dilemma: “justice too long delayed is justice denied”.

Baldwin, Eassey, and Brooke (2020)\citep{baldwin} further emphasize that U.S. courts faced “an existential threat more akin to a natural disaster,” forcing abrupt shifts to virtual operations while struggling to preserve fairness and procedural rights. These operational adaptations frame Cook County’s pandemic-era experience as part of a nationwide challenge to maintain continuity of justice amid crisis.

Fundamental principles of penal justice emphasize that morally similar cases should receive consistent punishments, regardless of external factors. However, recent research by Kundro et al. \citep{Kundro2023} revealed that time delays between crime and arrest significantly influence punishment severity. Their study combined controlled experiments with 6,029 participants, survey-based sampling, and archival analysis of sentencing data (160,772 civilian cases and police misconduct records), establishing that longer delays were associated with harsher punishments due to intensified perceptions of unfairness.

Our research builds on and extends this work by re-analyzing real-world judicial decisions using criminal case data from Cook County (2020–2024) \citep{CookCounty2024}, a period marked by systemic disruptions caused by the COVID-19 pandemic. According to the Washington Post report \citep{Witte2021}, "The question remains: is COVID solely to blame for the crisis that has overtaken the court system long after COVID caused the court shutdown?" This observation emphasizes the need to examine fundamental inefficiencies within judicial processes that were magnified during the pandemic.

While Kundro demonstrated the moral and perceptual consequences of time delays, our study focuses on how such delays manifest in real sentencing outcomes through distinct empirical methodologies. By bridging moral perception with judicial behavior, this study extends the existing literature on temporal justice and court efficiency, offering actionable insights into systemic disparities and the mechanisms through which time delays exacerbate punishment severity. We pursue the following objectives:

\begin{enumerate}
\item Validate and quantify the relationship between time delays and punishment severity in contemporary court decisions
\item Uncover natural patterns in sentencing outcomes through advanced clustering techniques, particularly focusing on how different case types respond to delays
\item Identify and rank the relative importance of factors influencing sentencing severity, including temporal, demographic, and case-specific variables
\end{enumerate}

\section{Data}
\vspace{-5pt} %
\subsection{Data Source and Phasing}

Our study utilizes administrative records from the Cook County State’s Attorney’s Office, consistent with prior empirical literature, and analyzes criminal cases from January 2020 to October 2024. To contextualize the timing of judicial disruptions, the study period is divided into three phases—Pre-COVID (January 2020–March 2020), Peak COVID (April 2020–December 2021), and Post-COVID (January 2022–October 2024)—based on milestones documented in the World Health Organization’s (WHO) \citep{Who2020} official timeline of the global COVID-19 response. This periodization reflects the escalation, peak disruption, and subsequent stabilization of institutional operations associated with the pandemic, allowing us to assess how judicial decision-making varied across distinct stages of systemic stress.

This phased framework aligns with scholarship suggesting that the COVID-19 pandemic acted not as a singular shock, but as a catalyst that intensified long-standing institutional inefficiencies. In particular, Godfrey, Richardson, and Walklate \citep{Godfrey2022} argue that the pandemic primarily amplified pre-existing court backlogs rather than creating fundamentally new challenges. Their findings support our interpretation of the pandemic era as an exogenous stressor that exposed structural vulnerabilities within the Cook County judicial system.

\subsection{Sample Selection and Variable Construction}
From an initial dataset of 303,963 records, we obtained 41,986 cases after restricting the sample to the defined study period. While approximately 78.3\% of these cases involved immediate arrests, our analysis focuses on the remaining 9,111 cases that experienced non-zero delays between offense and arrest. This restriction enables a more precise examination of how variation in procedural delay, defined as the temporal gap between crime and arrest, is associated with sentencing outcomes in cases where such delays are substantively meaningful.

The primary outcome variable, \textit{Weighted Punishment}, is a composite measure defined as the product of standardized sentence length, expressed in months, and an ordinal severity level ranging from 1 to 3 based on commitment type. This construction captures both the duration and the qualitative intensity of punishment within a single metric, facilitating comparison across sentencing modalities. Table 1 presents the complete set of variables used in the analysis.

\renewcommand{\arraystretch}{1.2}
\begin{longtable}{p{5.5cm} p{11.5cm}}
\caption{Dataset Features Description} \label{tab:features_description} \\
\toprule
\textbf{Variable Name} & \textbf{Description} \\
\midrule
\endfirsthead

\textit{Original Features} & \\
\midrule
Incident begin date & Date when the criminal incident occurred \\
Arrest date & Date when the defendant was arrested \\
Commitment term & Numerical value of sentence length \\
Commitment unit & Unit of measurement for sentence (e.g., months, years) \\
Commitment type & Type of sentence imposed (e.g., incarceration, probation) \\
Case length & Duration from case initiation to resolution in days \\
Age at incident & Defendant's age when crime occurred \\
Race & Defendant's racial identification \\
Gender & Defendant's gender identification \\
Updated offense category & Original detailed crime category \\
Sentence Judge & Identifier for presiding judge, serving as fixed court effect \\
Charge count & Number of charges filed in the case \\
\midrule
\textit{Derived Features} & \\
\midrule
Delay in days & Time difference between the incident date and arrest date, representing the length of delays\\
Standardized sentence months & All sentences converted into months to ensure consistency in punishment duration analysis \\
Crime group & Mapped offense categories into six groups: Violent, Property, Drug-Related, Sexual and Exploitation, Public Order and Regulatory, and Miscellaneous Crimes\\
COVID period & Categorical variable marking three pandemic phases: Pre-COVID (Jan 2020 – Mar 2020), Peak-COVID (Apr 2020 – Dec 2021), and Post-COVID (Jan 2022 – Oct 2024)\\
Case complexity & Interaction term between case length (in days) and charge count, representing administrative and legal complexity \\
Severity level & Scale 1–3 based on commitment type severity: incarceration, probation, and supervision or lighter sentences\\
\textbf{Weighted punishment}\newline (\textit{Target feature}) &Composite measurement: Sentence Months × Severity Level (1–3), capturing both duration and severity of punishment\\
\bottomrule
\end{longtable}

\newpage
\section{Methodology}
\vspace{-5pt} %
To examine the relationship between time delays and punishment severity, we employ a triangulated computational framework. This multi-method approach is motivated by the fact that judicial decision-making reflects both linear associations and more complex, non-linear interactions across cases.

We first use OLS regression to establish a baseline statistical relationship between procedural delay and sentencing severity, estimating how additional days of delay are associated with punishment outcomes while controlling for judge fixed effects and other covariates. To move beyond average effects and identify heterogeneity across cases, we then apply Bi-Secting K-Means clustering to uncover latent groupings in sentencing patterns. Finally, we implement a Random Forest model to address the high-dimensional nature of the data and to assess the relative importance of temporal variables alongside demographic and case-specific factors. Together, these methods provide complementary perspectives on the role of procedural delay in judicial outcomes.

\vspace{-5pt} %
\subsection{OLS Regression Model}

The OLS regression model was employed to examine whether the previously established relationship between time delays and sentencing severity remains robust in the context of Cook County data. The model quantifies the linear association between delay length and punishment severity while controlling for potential confounding factors. Interaction terms and fixed effects are included to account for systemic heterogeneity across courts and temporal variation.

Model performance was evaluated using multiple validation metrics. The R-squared value of 0.3595 indicates moderate explanatory power, with approximately 36\% of the variation in punishment severity explained by the included predictors. Information criteria (AIC = 26,437.68; BIC = 27,863.71) were used to assess the trade-off between model fit and complexity. The relatively high values reflect the model’s dimensionality and suggest scope for further refinement. While the linear specification cannot fully capture the complexity of sentencing decisions, it provides a transparent baseline for subsequent analyses.

Consistent with our findings, Godfrey et al.~\citep{Godfrey2022} document that resource scarcity and uneven digital adaptation during the pandemic deepened disparities in case processing and outcomes across jurisdictions. This broader pattern supports the interpretation that the association between procedural delay and sentencing severity in Cook County reflects structural constraints rather than isolated case dynamics.

\subsection{Bi-Secting K-Means Clustering}

To identify latent patterns in sentencing outcomes, we employ Bi-Secting K-Means clustering, a hierarchical variant of the K-Means algorithm that iteratively partitions the data into increasingly refined clusters. This approach is well suited to large, high-dimensional administrative datasets, as it balances computational scalability with interpretability while reducing sensitivity to initialization relative to standard K-Means. By allowing local optimization at each split, Bi-Secting K-Means can produce more stable and cohesive groupings than a single global partition, without the computational burden of fully hierarchical or model-based clustering methods.

\begin{figure}[ht]
\centering
\includegraphics[width=15cm]{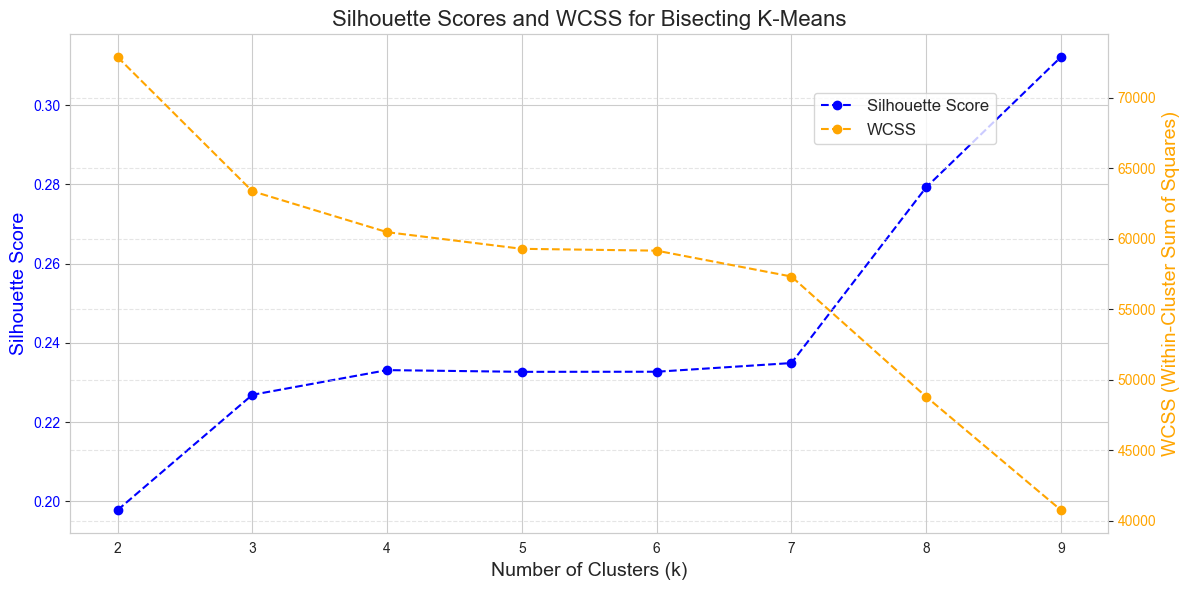}
\caption{Model Validation: Silhouette Scores and WCSS for Bisecting K-Means}
\label{fig:silhouette}
\end{figure}

Cluster validity was assessed using silhouette scores and within-cluster sum of squares (WCSS), as illustrated in Figure~\ref{fig:silhouette}. Silhouette scores peak at \(k = 4\), indicating the strongest balance between within-cluster cohesion and between-cluster separation at this value. Beyond \(k = 4\), improvements in silhouette scores are marginal and become unstable, while the WCSS curve exhibits diminishing returns, suggesting limited gains from additional clusters.

To further assess interpretability, we examine parallel coordinate plots for the four-cluster solution, which reveal clear separation along key dimensions including delay length, case complexity, and sentencing severity. Together, these diagnostics indicate that \(k = 4\) provides a parsimonious representation of heterogeneity in sentencing outcomes without introducing unnecessary fragmentation.

\subsection{Random Forest Model}

The Random Forest model was selected for its ability to accommodate non-linear relationships, handle high-dimensional data, and provide interpretable measures of feature importance without requiring feature scaling. This ensemble approach is well suited to complex administrative datasets in which interactions among temporal, demographic, and case-specific variables may not be well captured by linear specifications.

Model implementation involved hyperparameter tuning via grid search with 5-fold cross-validation to balance bias and variance. Predictive performance was evaluated using standard classification metrics, including accuracy, precision, and recall. While predictive accuracy is not the primary objective of analysis, the Random Forest model is used to assess the relative importance of procedural delay alongside other covariates, offering complementary insight into factors most strongly associated with sentencing severity.

\vspace{-5pt} %

\section{Results}
\vspace{-5pt} %
\subsection{OLS Regression Model}

\begin{table}[htbp]
\centering
\caption{OLS Regression Results}
\label{tab:ols_results}
\begin{tabular}{lrrrr}
\toprule
\textbf{Variable} & \textbf{Coefficient} & \textbf{Std.Error} & \textbf{t-statistic} & $\mathbf{P > |t|}$\\
\midrule
Intercept        & 0.2578  & 0.0190 & 13.589 & 0.0000 \\
Delay in days    & 0.0196  & 0.0070 & 2.675  & 0.0070 \\
Charge counts    & -0.0462 & 0.0090 & -4.926 & 0.0000 \\
Length of cases  & 0.0007  & 0.0001 & 10.118 & 0.0000 \\
Case complexity  & 0.0001  & 0.0000 & 4.265  & 0.0000 \\
Age at incident  & 0.0030  & 0.0010 & 2.946  & 0.0030 \\
COVID Period     & 0.1613  & 0.0550 & 2.945  & 0.0030 \\
Race: Black      & 0.1885  & 0.1030 & 1.834  & 0.0670 \\
Gender: Male     & 0.2485  & 0.0340 & 7.244  & 0.0000 \\
\bottomrule
\end{tabular}
\end{table}

The OLS regression results indicate a positive and statistically significant association between procedural delay and punishment severity after controlling for key covariates. Longer delays are associated with harsher sentences, although the estimated effect size is relatively modest, with each additional day of delay corresponding to an increase of 0.0196 units in the weighted punishment measure. Demographic characteristics, including male gender and older age at the time of the incident, are also associated with higher sentencing severity. Race, measured as an indicator for Black defendants, is not statistically significant at the 5\% level, though the near-threshold \textit{p}-value suggests a potential association that warrants further examination.

The COVID-19 period is associated with significantly harsher sentencing outcomes, consistent with sustained systemic disruptions such as court closures, procedural backlogs, and altered judicial operations during the pandemic. The model explains approximately 36\% of the variation in sentencing severity, indicating that additional unobserved factors, including judicial discretion and courtroom-level dynamics, likely contribute to sentencing outcomes.

Supplementary analyses by offense category reveal particularly severe sentencing patterns for violent offenses, sexual exploitation crimes, and certain property-related offenses, including aggravated robbery, homicide, sex crimes, human trafficking, home invasion, and vehicular hijacking. These patterns suggest that the association between procedural delay and sentencing severity may be more pronounced in cases involving serious offenses.

Overall, the findings are consistent with prior literature documenting a positive association between procedural delay and punishment severity. At the same time, the relatively modest magnitude of the estimated delay effect highlights the importance of exploring alternative model specifications, including non-linear relationships, to better capture the complexity of sentencing decisions.

\subsection{Bi-Secting K-means Clustering}
\begin{figure}
        \centering
        \includegraphics[width=0.9\linewidth]{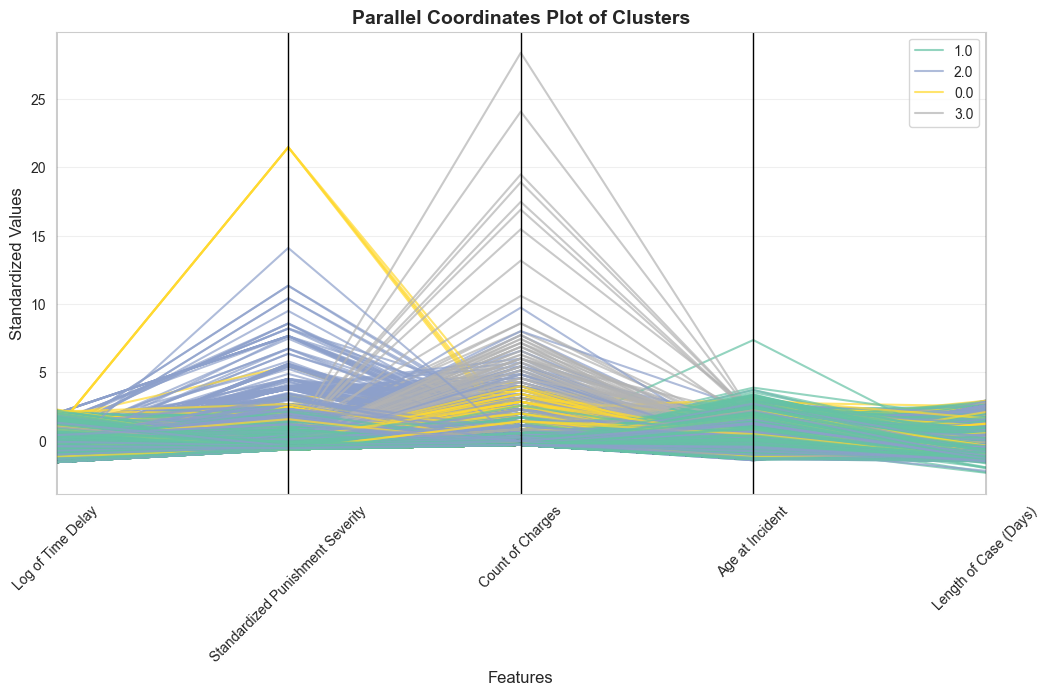}
        \caption{Parallel Coordinate Plot of Clusters}
        \label{fig:parellel}
    \end{figure}

The clustering analysis identifies four distinct groups of sentencing outcomes, illustrating how procedural delay, case complexity, and offense characteristics jointly shape punishment severity:

\begin{itemize}
    \item \textbf{Cluster 0}: Cases involving sexual and exploitation-related offenses with high punishment severity, characterized by longer delays and extended case resolution times (528 cases).
    \item \textbf{Cluster 1}: Lower-severity cases, predominantly involving property crimes, with shorter delays and lighter punishments (4,531 cases).
    \item \textbf{Cluster 2}: Serious cases involving violent offenses, such as aggravated robbery and homicide, associated with longer delays and elevated punishment severity (2,594 cases).
    \item \textbf{Cluster 3}: A small outlier group of highly complex cases with exceptionally high charge counts, resulting in substantial delays and severe punishments (51 cases).
\end{itemize}

The parallel coordinates plot in Figure~\ref{fig:parellel} shows that Cluster 0 exhibits the highest levels of punishment severity, followed by Cluster 2, while Cluster 1 represents the most common and comparatively less severe cases. Cluster 3 displays pronounced heterogeneity across charge counts and delay length, reflecting its outlier status.

These patterns suggest that procedural delay and case complexity play a particularly important role in sentencing outcomes for serious offenses, including sexual exploitation and violent crimes. The distinct profile of Cluster 3 further highlights the challenges posed by highly complex cases, which may require specialized administrative or judicial handling. Overall, the clustering results reinforce the conclusion that temporal delay and institutional complexity are central dimensions of heterogeneity in sentencing outcomes and underscore the relevance of targeted reforms aimed at reducing delay-related disparities.

Taken together, these cluster-level patterns complement the regression and machine learning results by illustrating how the association between procedural delay and sentencing severity varies systematically across distinct case profiles.

\subsection{Random Forest Model}

The Random Forest model was fine-tuned using grid search with five-fold cross-validation, evaluating 300 hyperparameter combinations across 1,500 model fits. The selected specification uses bootstrap sampling, a maximum tree depth of 34, log$_2$ feature sampling at each split, a minimum of one observation per leaf node, a minimum of four observations to split an internal node, and 151 trees. The optimized model achieved an overall classification accuracy of 71\%, with balanced precision (0.70–0.73) and recall (0.70–0.71). While these metrics indicate stable performance across classes, the overall accuracy remains moderate, reflecting the complexity of sentencing decisions and the high-dimensional structure of the data.

\begin{figure}
    \centering
    \includegraphics[width=1\linewidth]{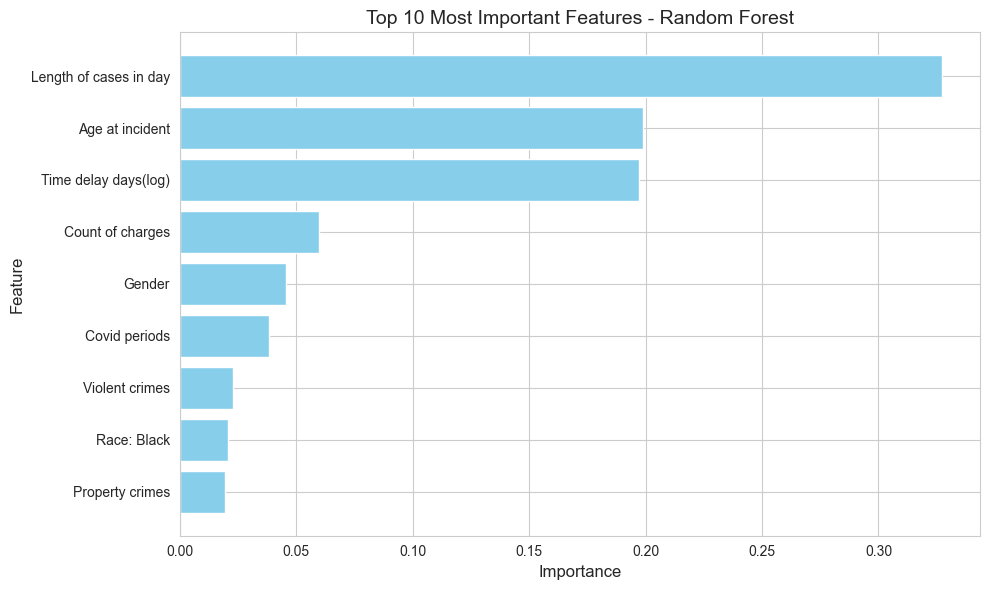}
    \caption{Random Forest Feature Importance}
    \label{fig:outputimportance}
\end{figure}

Beyond predictive performance, the Random Forest model is used to assess the relative importance of features associated with punishment severity. As shown in Figure~\ref{fig:outputimportance}, the most influential predictors include the length of the case in days, age at incident, and the log of time delay, followed by charge count and other demographic and offense-related variables.

\begin{itemize}
    \item \textbf{Length of case in days} emerges as the strongest predictor, reflecting the administrative and legal complexity associated with more severe cases.
    \item \textbf{Age at incident} is positively associated with sentencing severity.
    \item \textbf{Procedural delay} remains a prominent predictor, reinforcing the role of temporal factors in shaping sentencing outcomes.
    \item Additional variables, including \textbf{charge count}, \textbf{gender}, \textbf{COVID period}, and \textbf{crime type}, also contribute to sentencing severity, consistent with patterns observed in the OLS regression analysis.
\end{itemize}

\section{Conclusions and Implications}
\vspace{-5pt} %

\subsection{Conclusions}

This study demonstrates that procedural delays in the criminal justice system are not merely administrative artifacts but are meaningfully associated with sentencing severity. Using merged administrative data from Cook County (2020–2024), the analysis shows that longer delays between offense and arrest are consistently associated with harsher punishments, even after accounting for a wide range of demographic, temporal, and case-specific factors. Across regression, clustering, and machine-learning approaches, delay-related measures retain explanatory and predictive relevance, suggesting that temporal frictions systematically shape judicial outcomes.

These findings contribute to a broader understanding of how institutional processes influence the administration of justice. The consistency of delay-related associations across methods indicates that sentencing outcomes are sensitive to features of court operations that fall outside the formal legal merits of a case. The results also highlight the role of administrative capacity, docket management, and information availability in shaping judicial decision-making. Procedural delays appear to function not only as markers of case complexity or system congestion but also as institutional conditions under which harsher penalties are more likely to arise.

Although this study does not establish causal relationships, it provides empirical evidence consistent with emerging theoretical perspectives on judicial behavior, particularly the view that decision-makers respond to contextual pressures, incomplete information, and workload constraints. By documenting systematic associations rather than isolated anomalies, the findings underscore the importance of viewing procedural delay as an equity concern with measurable distributional consequences rather than as a neutral administrative feature of adjudication.

Taken together, the results support a broader conclusion: timely and efficient case processing is not only a matter of administrative performance but also a central dimension of fairness in sentencing. Future work drawing on quasi-experimental designs and multi-jurisdictional data can further strengthen the evidence base and help guide reforms aimed at reducing delay-driven disparities in criminal justice.

\subsection{Policy and Research Implications}

This study extends the findings of Kundro et al. \citep{Kundro2023} by translating experimental insights on temporal justice into an empirical analysis of real-world judicial behavior. By combining regression, clustering, and machine learning approaches, the results show that time delays between offense and arrest are systematically associated with harsher sentencing outcomes, even after accounting for demographic, procedural, and temporal factors. These findings provide robust empirical support for the argument that institutional inefficiencies, particularly those related to time and administrative capacity, have tangible consequences for justice and equity. In doing so, the analysis complements recent administrative-data evidence demonstrating that court delays have substantive behavioral and institutional effects rather than serving as neutral features of case processing \citep{andersen2024}.

Beyond validating the effect of temporal delays, the analysis reveals broader systemic disparities in the administration of justice. Sentencing outcomes are influenced not only by legal parameters but also by the social characteristics of defendants and the structural constraints under which courts operate. Demographic effects, such as the harsher sentencing of male and older defendants, reflect underlying inequities in the exercise of judicial discretion, while the near-significance of race highlights the continued importance of social identity in shaping legal outcomes. These patterns are consistent with scholarship emphasizing that delay is not merely incidental but embedded within complex court systems, where institutional routines and path-dependent practices can systematically reproduce inequality \citep{albrecht2022}.

The COVID-19 pandemic further magnified these inequities and exposed the fragility of institutional systems under stress. As court operations slowed, procedural backlogs expanded, compounding the effects of pre-existing disparities. Disruptions to hearings and altered procedural norms contributed to sentencing variability and intensified punitive outcomes. These disruptions underscore how administrative strain can alter the conditions under which judicial discretion is exercised, reinforcing concerns that delay functions as an additional and unevenly distributed burden within the criminal justice system \citep{albrecht2022}.

However, rapid digital adaptation has created a new set of challenges. Many jurisdictions turned to virtual proceedings as a stopgap solution, but the uneven rollout of technology risks deepening inequities rather than resolving them. As Baldwin et al. \citep{baldwin} caution, early virtual-court innovations often reproduced existing inequities due to inconsistent access to technology and limited cybersecurity guidance. Addressing these digital divides is thus essential for any long-term strategy to reduce delay and strengthen judicial legitimacy.

From a policy perspective, the findings emphasize the importance of judicial and administrative reforms to mitigate procedural delays, reduce case backlogs, and address demographic biases that undermine equitable outcomes. Evidence from both U.S. and international contexts suggests that delays can shape legal outcomes and post-adjudication behavior through multiple channels, including stress, uncertainty, and altered incentives. Integrating data-driven governance tools, such as predictive analytics and transparency dashboards, can support real-time monitoring of sentencing disparities and resource allocation. Ultimately, this study situates sentencing practices within a broader discussion of how institutional processes shape inequality and well-being. Justice reform should therefore be viewed not merely as a procedural concern but as a foundational component of social equity. By advancing empirical methods in computational legal analytics, this work contributes to a more rigorous and data-informed understanding of justice, fairness, and institutional performance. Future work could extend this framework across jurisdictions and apply causal inference methods to isolate the specific effects of administrative delays on sentencing outcomes.

\section{Limitations}
\vspace{-5pt} %

Our analysis is not as methodologically robust as the original study by Kundro et al. \citep{Kundro2023}, which draws on six experiments combining controlled laboratory designs, survey-based moral perception measures, and archival analyses across multiple datasets. In contrast, this study focuses exclusively on administrative criminal case data from Cook County between 2020 and 2024, which may limit the generalizability of findings to other jurisdictions or to national sentencing patterns. Whereas the original literature explicitly examines public perceptions and moral judgments, our approach shifts the analytical focus from attitudinal mechanisms to institutional and case-specific factors embedded in real-world judicial decision-making. This divergence yields complementary insights but also complicates direct comparisons across empirical settings.

A key limitation lies in the restricted variability of procedural delays within the dataset. Approximately 80\% of cases involve arrests occurring within a single day of the offense, resulting in limited dispersion along the temporal dimension of interest. This concentration constrains the ability to observe the full range of behavioral and institutional dynamics theorized in prior work. While Kundro et al. \citep{Kundro2023} posit that longer delays may intensify punishment severity through mechanisms such as moral outrage and deterrence, the predominance of short delays in this sample limits empirical leverage to directly test these psychological pathways. More broadly, recent syntheses emphasize that identifying the behavioral and deterrent effects of punishment timing is empirically challenging in observational settings, particularly when variation in timing is limited or endogenous to institutional processes \citep{LoefflerNagin2022}. This constraint highlights the value of future research designs incorporating richer temporal heterogeneity, quasi-experimental variation, or complementary survey-based evidence.

The high-dimensional structure of the dataset further presents analytical challenges. With numerous categorical variables and derived features, clustering methods such as Bi-Secting K-Means may perform suboptimally, as they are generally better suited to lower-dimensional feature spaces. Similarly, the OLS regression framework imposes linearity assumptions that may fail to capture non-linear interactions or threshold effects characteristic of complex judicial processes. Although the Random Forest model partially relaxes these constraints, the inclusion of over 200 predictors may introduce noise and reduce out-of-sample generalizability. Future analyses could benefit from dimensionality-reduction or regularization techniques, such as principal component analysis (PCA), LASSO, or Elastic Net, to improve model interpretability and robustness.

Finally, although the inclusion of judge fixed effects addresses some forms of unobserved heterogeneity, judicial discretion and courtroom-level dynamics may operate through more deeply nested institutional structures. Prior research suggests that procedural delay is often embedded within routinized court practices and administrative constraints rather than arising solely from individual case characteristics \citep{albrecht2022}. Multi-level or hierarchical modeling approaches may therefore be better suited to disentangling variation across judges, court divisions, and time periods. Additionally, while the COVID-19 period is incorporated as a categorical control, broader pandemic-era disruptions—such as staffing shortages, shifting prosecutorial priorities, and evolving procedural norms—introduce unobserved confounders beyond the scope of this analysis. Addressing these limitations through multi-jurisdictional data integration and mixed-methods approaches would strengthen both the external validity and theoretical precision of future research.

\end{document}